\documentclass[amsmath, amssymb, preprintnumbers, showpacs, showkeys,aps,prl,superscriptaddress,twocolumn]{revtex4-2}

\usepackage{amsmath}
\usepackage{amssymb}
\usepackage{gensymb}
\usepackage{bm}
\usepackage{graphicx}
\usepackage{float}
\usepackage[dvipsnames]{xcolor}
\usepackage{placeins}
\usepackage{braket}
\usepackage{bbold}
\usepackage{ulem}
\usepackage[colorlinks, linkcolor=blue, citecolor=blue, urlcolor=blue, breaklinks=red]{hyperref}

\newcommand{\I}{{\rm Im\,}}
\newcommand{\R}{{\rm Re\,}}
\newcommand{\ketbra}[2]{|#1\rangle\langle #2|}
\newcommand{\SubFig}[2]{\ref{#1}{\color{blue}#2}}
\usepackage{bbm}
\newcommand{\1}{\mathbbm{1}}

\usepackage{dsfont}
\usepackage{orcidlink}

\newcommand{\UFSCar}{Departamento de Física, Universidade Federal de São Carlos, \\Rodovia Washington Luís, km 235 - SP-310, 13565-905 São Carlos, SP, Brazil}
\newcommand{\Nice}{Universit\'e C\^ote d'Azur, CNRS, Institut de Physique de Nice, 06560 Valbonne, France}
\newcommand{\CSIC}{Instituto de Física Fundamental (IFF), Consejo Superior de Investigaciones Científicas (CSIC), \\ Calle Serrano 113b, 28006 Madrid, Spain}

\definecolor{Mycolor1}{HTML}{44aa99}
\definecolor{Mycolor2}{HTML}{cc6677}
\begin{document}
 
\title{Detecting entanglement from macroscopic measurements of the electric field and its fluctuations}

\author{Pedro Rosario~\orcidlink{0000-0002-7628-7373}}
\email{pedrorosario@estudante.ufscar.br}
\affiliation{\UFSCar}

\author{Alan C. Santos~\orcidlink{0000-0002-6989-7958}}
\email{ac\_santos@iff.csic.es}
\affiliation{\UFSCar}
\affiliation{\CSIC}

\author{Nicola Piovella}
\email{nicola.piovella@mi.infn.it}
\affiliation{Dipartimento di Fisica ``Aldo Pontremoli'', Universit\`a degli Studi di Milano, Via Celoria 16, I-20133 Milano, Italy \& INFN Sezione di Milano, Via Celoria 16, I-20133 Milano, Italy}

\author{Robin Kaiser}
\email{robin.kaiser@univ-cotedazur.fr}
\affiliation{\Nice}

\author{André Cidrim~\orcidlink{0000-0003-0007-2330}}
\email{andrecidrim@gmail.com}
\affiliation{\UFSCar}

\author{Romain Bachelard~\orcidlink{0000-0002-6026-509X}}
\email{romain@ufscar.br}
\affiliation{\UFSCar}

\begin{abstract}
To address the outstanding task of detecting entanglement in large quantum systems, entanglement witnesses have emerged, addressing the separable nature of a state. Yet optimizing witnesses, or accessing them experimentally, often remains a challenge. We here introduce a family of entanglement witnesses for open quantum systems, based on the electric field --- its quadratures and the total fluorescence. More general than spin-squeezing inequalities, it can detect new classes of entangled states, as changing the direction for far-field observation opens up a continuous family of witnesses, without the need for a state tomography. Their efficiency is demonstrated by detecting, from almost any direction, the entanglement of collective single-photon states, such as long-lived states generated by cooperative spontaneous emission. Able to detect entanglement in large quantum systems, these electric-field-based witnesses can be used on any set of emitters described by the Pauli group, such as atomic systems (cold atoms and trapped ions), giant atoms, color centers, and superconducting qubits.
\end{abstract}
	
\date{\today}

\maketitle

{\it Introduction.---}The detection of multipartite entanglement remains a challenging problem, due to the difficulty of realizing state tomography and the lack of tools to quantify it once the system state is known~\cite{Bennett_1996, Hill_1997, Wootters_98,Vidal_2002,G_hne_2009}. For low dimensional systems, for instance, entanglement witnesses such as the positive partial transpose of the density matrix~\cite{Peres_1996}, the majorization criterion~\cite{Nielsen_2001}, and many other proposals~\cite{Horodecki_1997, Hofmann_2003, Shchukin_2005, Horodecki_2006}, require state tomography. When the density matrix is not an available resource (and this is usually the case for quantum systems with a large Hilbert space), a different approach to the entanglement detection problem needs to be considered, typically relying on the direct measurement of quantum observables $\hat{\mathcal{O}}$~\cite{Terhal_2000,Bru_2002,Horodecki_1996_1}. More specifically, if $\hat{\rho}$ represents a quantum state, then $\hat{\mathcal{O}}$ is an entanglement witness if and only if a violation of the inequality $\text{Tr}(\hat{\mathcal{O}}\hat{\rho})\geq 0$ implies an entangled nature of state $\hat{\rho}$. This approach has put forward the investigation of multipartite entanglement in many-particle states~\cite{2013_Sperling_Vogel_prl, 2018_Friis_Huber_natrevphys, Friis2018,2024_Sun_Yu_prl}. In spin-like systems where the statistics (first and second moments) of a collective spin operator can be accessed experimentally, several sets of inequalities detecting spin squeezing have been proposed as entanglement witnesses~\cite{S_rensen_2001,Korbicz_2005,Toth_2007,Toth_2009}. These inequalities have, in turn, stimulated the definition of ``metrologically-useful'' squeezed states~\cite{2018_Pezze_Treutlein_rmp,Bornet_2023}, which enable reducing uncertainties in interferometry measurements beyond the so-called standard quantum limit~\cite{2011_Ma_Nori_physrep}. To access larger classes of entangled states in systems where local measurements are particularly challenging, other entanglement witnesses taking into account the distance between particles have been proposed~\cite{Krammer_2009,Lee_2019}. However, identifying experimentally accessible or more optimal witnesses remains a largely open challenge.

To address this problem, we introduce a continuous set of inequalities based on the measurement of the electric field --- the quadratures, the total fluorescence, and their fluctuations, more precisely. While they reduce to spin-squeezing inequalities in particular geometries and observation angles, changing the direction of detection of the field allows one to probe an infinity of entanglement witnesses, as the optical path from the different atoms to the detection changes. This family of witnesses is thus more optimal than spin-squeezing ones and applies to quantum emitters described by the Pauli group --- from natural two-level atoms to artificial ones with inhomogeneous broadening. 


{\it Entanglement witnesses from the electric field.---}Let us consider an ensemble of $N$ two-level emitters, with $\ket{\uparrow}_j$ and $\ket{\downarrow}_j$ the excited and ground states for each atom $j$. Without local access to the emitters to realize state tomography, collective information on the system state can still be extracted from the radiated electric field. While the operator and the study of its momenta (intensity and field fluctuations) is a keystone of quantum optics~\cite{Glauber1963,Glauber1963b,Loudon:book}, we here show that it provides precious information on the atomic system state as well. In the far-field limit, the electric field operator for the two-level atoms reads
\begin{align}
\hat{E}_\mathbf{k}^{\pm}=\sum_{j=1}^{N}e^{\mp i\mathbf{k}.\mathbf{r}_j}\hat{\sigma}^{\mp}_{j},
    \label{eq:far-field}
\end{align}
with $\mathbf{k}$ the direction of observation, $\hat{\sigma}^{+}_{j}=\ket{\uparrow}\bra{\downarrow}_j$($\hat{\sigma}^{-}_{j}=\ket{\downarrow}\bra{\uparrow}_j$) the raising (lowering) Pauli operators of atom $j$, and $\mathbf{r}_j$ its position. Without loss of generality, the prefactor in front of the electric field operator has been set to unity. We then introduce the field quadratures $\hat{X}_\mathbf{k}$ and $\hat{Y}_\mathbf{k}$, and inversion population operator $\hat{Z}$,
\begin{subequations}
\label{eq:Operators}
\begin{align}
&\hat{X}_\mathbf{k}=\hat{E}_\mathbf{k}^{+}+\hat{E}_\mathbf{k}^{-},
\\    & \hat{Y}_\mathbf{k}=i(\hat{E}_\mathbf{k}^{+}-\hat{E}_\mathbf{k}^{-}), 
\\    &\hat{Z}=\sum_{j=1}^{N}\hat{\sigma}^{z}_{j},
\end{align}
\end{subequations}
with $\hat{\sigma}^{z}_{j}=\ket{\uparrow}\bra{\uparrow}_j-\ket{\downarrow}\bra{\downarrow}_j$ the inversion population operator for atom $j$.

Any arbitrary separable state of $N$ particles can be written as $\hat{\rho}=\sum_{l=1}^{L}p_{l}\hat\rho^{(l)}_{1}\otimes \hat\rho^{(l)}_{2}\otimes \hdots \otimes \hat\rho^{(l)}_{N}$ with $\sum_{l=1}^{L}p_{l}=1$ \cite{Werner_1989}, where the superscript $(l)$ denotes a local state of the statistical mixture. If the state cannot be written in this form, there is thus entanglement between at least two particles. Following this statement, we now introduce our electric-field-based witness for entanglement~\cite{SM}.

\textbf{Theorem: } If there exists a wavevector $\mathbf{k}$ for which a quantum state $\hat{\rho}$ satisfies the inequality 
\begin{align}
    W_\mathbf{k}=\min\left\{w_{1,\mathbf{k}}, w_{2,\mathbf{k}}, w^{\alpha,\beta,\gamma}_{3,\mathbf{k}}, w^{\alpha,\beta,\gamma}_{4,\mathbf{k}}\right\} < 0,
    \label{eq:Total_Ineq}
\end{align}
then $\hat{\rho}$ is an entangled state. 

The witness $W_\mathbf{k}$ encompasses the following series of entanglement witnesses
\begin{subequations}
\label{eq:Inequalities}
\begin{align}
  \label{ineq1}  w_{1,\mathbf{k}} & =  N(2+N)  - \langle \hat{X}^{2}_\mathbf{k}\rangle - \langle \hat{Y}^{2}_\mathbf{k}\rangle - \langle \hat{Z}^{2}\rangle,\\
    \label{ineq2} w_{2,\mathbf{k}} &= (\Delta \hat{X}_\mathbf{k})^{2}+(\Delta \hat{Y}_\mathbf{k})^{2}+(\Delta \hat{Z})^{2}  -2N,\\
    \label{ineq3} w^{\hat{A},\hat{B},\hat{C}}_{3,\mathbf{k}} &= 2N +(N-1)(\Delta \hat{A})^{2}-\langle \hat{B}^{2}\rangle -\langle \hat{C}^{2} \rangle ,\\  w^{\hat{A},\hat{B},\hat{C}}_{4,\mathbf{k}} &=(N-1)\left[(\Delta \hat{A})^{2}+(\Delta \hat{B})^{2}\right]-\langle \hat{C}^{2} \rangle \nonumber
    \\ &-N (N-2),
    \label{ineq4}
\end{align}
\end{subequations}
where $(\Delta \bullet )^{2}=\langle \bullet ^{2}\rangle - \langle \bullet \rangle ^{2}$ corresponds to the variance, and superscript $\{\hat{A},\hat{B},\hat{C}\}$ to the cyclic permutations over the set $\{\hat{X}_\mathbf{k},\hat{Y}_\mathbf{k},\hat{Z}\}$.

The derivation of the witness relies on showing that all separable states fulfill the inequalities $w_{n,\mathbf{k}}\geq 0$ for $n=1...4$ (using a concavity argument~\cite{SM}). Hence, a state satisfying $W_\mathbf{k}<0$ violates at least one of these inequalities and is thus entangled. 

While the argument is similar to the one used to derive spin-squeezing inequalities, we point out that our electric-field witnesses represent a much broader family of witnesses. More specifically, the original inequality proposed by S{\o}rensen \textit{et al.}~\cite{S_rensen_2001} was generalized to a finite set of inequalities by T\'oth \textit{et al.}~\cite{Toth_2007,Toth_2009}, to account for the different components of the collective spin operators. The relation between~\eqref{eq:Inequalities} and the spin squeezing-inequalities of Refs.~\cite{S_rensen_2001,Toth_2007,Toth_2009} is obtained by setting $\mathbf{k}=\mathbf{0}$, so they are hereafter denoted by $W_{\mathbf{0}}$. Differently, the present family of inequalities~\eqref{eq:Inequalities} is infinite, and one can span the witnesses by changing the direction of detection of the light. Changing the light wavenumber $k$ also provides a broader class of witness~\cite{Krammer_2009, Lee_2019}. We here focus on close-to-resonance witnesses. Indeed, far from resonance, one enters the fully dispersive regime, often used for quantum non-demolition measurements~\cite{Braginsky1996, Grangier1998}. Nevertheless, this limit is beyond the scope of this work.


\begin{figure}[t!]
\includegraphics[width=\columnwidth]{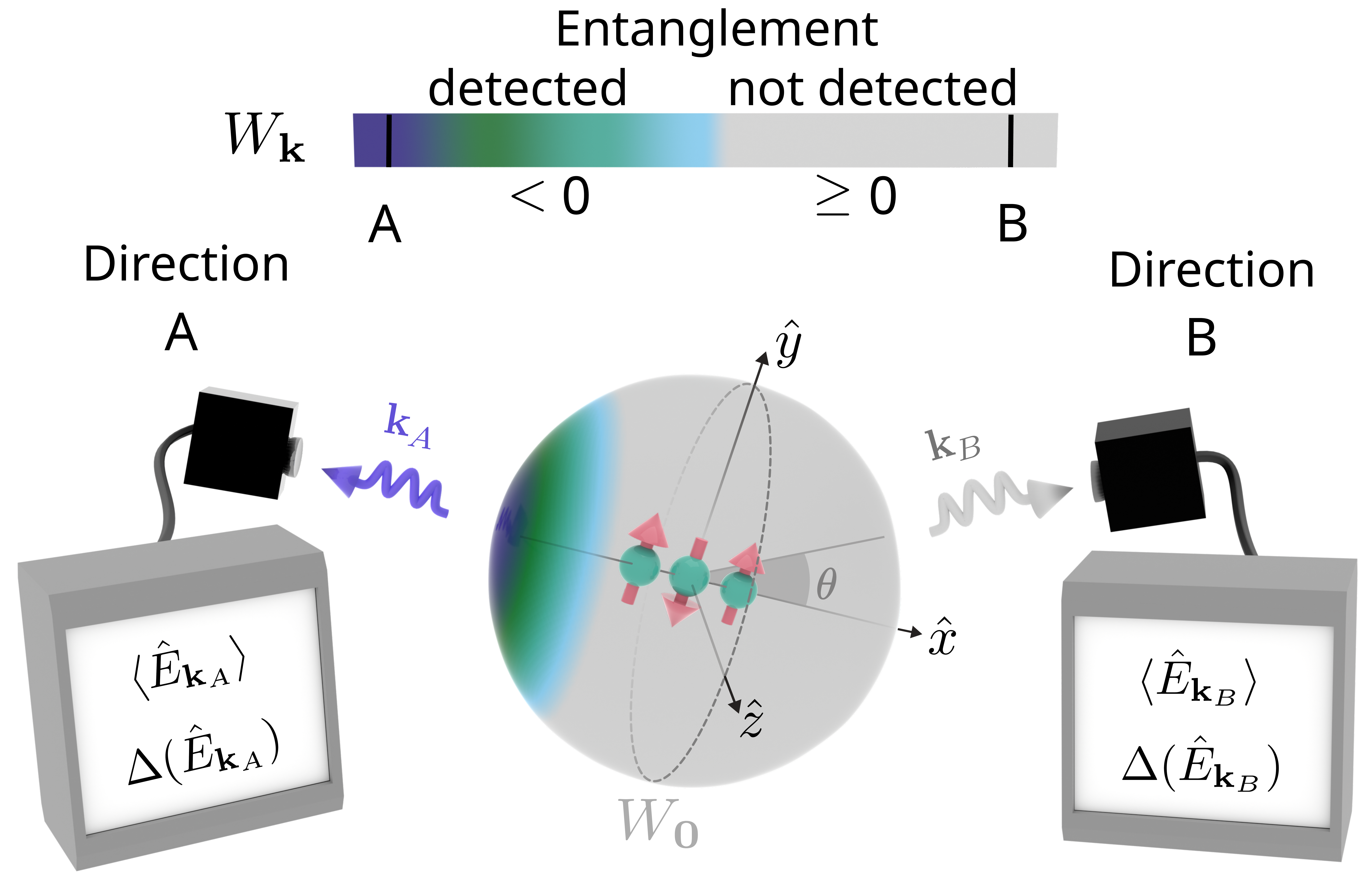}
\caption{System of three atoms in free space, prepared in the entangled state~\eqref{eq:main_state}. The sphere depicts the values of the entanglement witness $W_\mathbf{k}$ from Eq.~\eqref{eq:Inequalities}, monitored along different directions $\mathbf{k}$ through measurement of the field operators and their second moments. The colored area stands for directions where entanglement is detected ($W_{\mathbf{k}}<0$, here since either \eqref{ineq2} or \eqref{ineq3} is violated, such as in direction $A$), whereas the gray one corresponds to directions where it is not detected ($W_{\mathbf{k}}\geq 0$, e.g., position $B$). The dashed circumference on the sphere indicates the observation directions (orthogonal to the atomic chain) along which $W_{\mathbf{k}}=W_{\mathbf{0}}$.}
\label{Fig:Scheme}
\end{figure}

{\it Three-atom case.---}Let us now illustrate the advantage of electric-field-based inequalities by considering the light scattered by a linear arrangement of three atoms along the $\hat{x}$-axis, prepared in the following state
\begin{align}
    \ket{\psi}=\frac{1}{\sqrt{3}}\left(\ket{\uparrow\uparrow \downarrow}+e^{i\Lambda}\ket{\downarrow\uparrow\uparrow}+e^{i2\Lambda}\ket{\downarrow\downarrow\uparrow}\right),
    \label{eq:main_state}
\end{align}
equally spaced by $d=0.3/k$ and a relative phase $\Lambda=\pi/3$. Figure~\ref{Fig:Scheme} exhibits the behavior of the entanglement witness $W_\mathbf{k}$ for this system along different directions of observation $\mathbf{k}$, based on the measurement of the field operators~\eqref{eq:Operators} and their fluctuations. In this particular one-dimensional configuration, the spin-squeezing inequality $W_\mathbf{0}$ can be directly probed with the electric field in directions orthogonal to the chain (dashed circumference on the $yz$-plane in Fig.~\ref{Fig:Scheme}), since the optical path until the detector is the same for all atoms [$\mathbf{k}\cdot(\mathbf{r}_j-\mathbf{r}_m)\equiv 0$]. Note that $W_{\mathbf{0}}> 0$, meaning that spin-squeezing inequalities do not detect the entanglement of this state. Differently, the statistics of the scattered light can capture it in some directions of observation. The colored area corresponds to the directions where the field measurement allows for the detection of entanglement, and the gray-colored one to directions where $W_\mathbf{k}\geq0$, so the entanglement is not detected by witness~\eqref{eq:Total_Ineq}. This demonstrates how the degree of freedom of the phase terms in the electric field allows for a more optimal entanglement detection.


{\it Single-excitation states.---}We now consider an $N$-atom single-excitation Dicke state 
\begin{equation}
    \ket{\mathcal{D}_N}=\frac{1}{\sqrt{N}}\sum_{n=1}^N e^{i\phi_{n}}\ket{\uparrow_{n}},\label{eq:TDS}
\end{equation}
where $\ket{\uparrow_{n}}$ represents a state where atom $n$ is excited and all the others are in the ground state. These states are detected by electric field witnesses {\it in almost all directions}. Indeed, one can show that, for these states~\cite{SM}
\begin{equation}
w_{2,\mathbf{k}} =-w_{3,\mathbf{k}}^{\hat{Z},\hat{Y}_\mathbf{k},\hat{X}_\mathbf{k}}=S_\mathbf{k},
\label{eq:dsum}
\end{equation}
with $S_\mathbf{k}\equiv -\frac{4}{N}\sum_{j=1}^{N}\sum_{s\neq j}^{N}\cos{\left[\phi_{s}-\phi_{j}+\mathbf{k}\cdot(\mathbf{r}_{j}-\mathbf{r}_{s})\right]}$. A negative $S_\mathbf{k}$ violates inequality~\eqref{ineq2}, while a positive $S_\mathbf{k}$ violates inequality~\eqref{ineq3}. Hence, only the directions of observation $\mathbf{k}$ which satisfy $S_\mathbf{k}=0$ are inadequate to detect the entanglement from the field fluctuations: It is a set of measure zero. 

\begin{figure*}[t!]
\includegraphics[width=2\columnwidth]{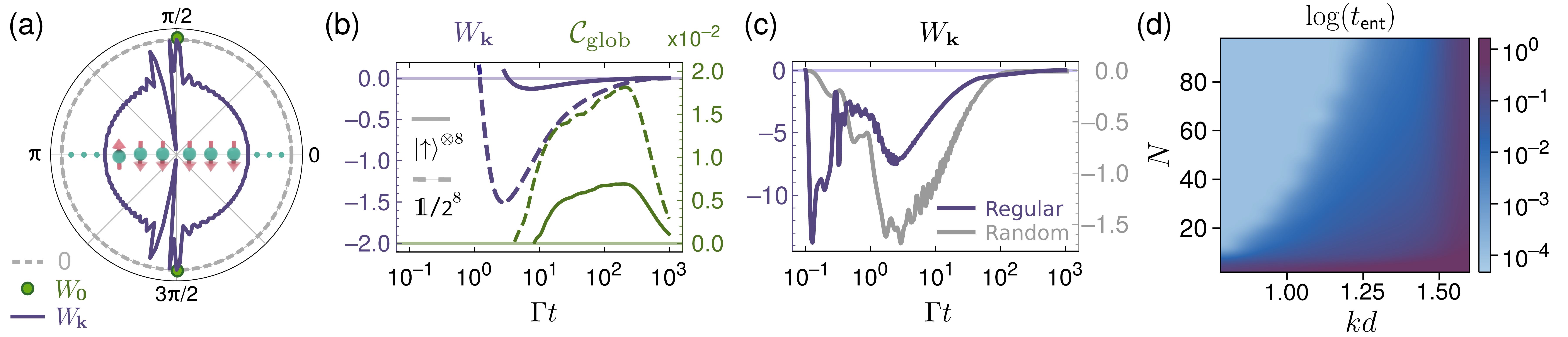}
\caption{(a) Witness $W_\mathbf{k}$ along different directions, for a one-dimensional chain of $N=100$ atoms in the single-excitation Dicke state~\eqref{eq:TDS}, with phases of the form $\phi_n = n\arccos\delta$ ($\delta\approx 0.997$ is chosen so the spin squeezing inequality is not violated, see text). (b) Evolution of the witness $W_\mathbf{k}$ (in blue) and of the concurrence $\mathcal{C}_{\text{glob}}$ (in green) during the decay dynamics when the system starts either in the fully excited state $\ket{\uparrow}^{\otimes 8}$ (plain curves) or in the fully classical mixed state $\hat{\rho}_{0}=\1/2^{8}$ (dashed curves). (c) Dynamics of the entanglement witness for a regular chain with spacing $d=0.3/k$ (gray curve) and a disordered three-dimensional spherical cloud of radius $2/k$ (blue curve)  of $N=8$ atoms, initially in the separable, antisymmetric state $\ket{\text{A}}$. The entanglement witness exhibits similar behavior in the two cases. (d) Time $t_\mathrm{ent}$ at which entanglement is first detected at an observation angle of $0.45\,\pi$ from the chain axis, as a function of the number of particles $N$ and lattice spacing $kd$. The witness $W_{\mathbf{k}}$ is calculated using a second-order cumulant expansion, with the system initially in state $\ket{\text{A}}$.}
  \label{Fig:Decay}
\end{figure*}

Let us illustrate this feature by considering a regular chain of $N=100$ atoms, again along the $\hat{x}$-axis, with spacing $d=\pi/2k$. Taking into account the rotational symmetry around the $\hat{x}$-axis, the light is monitored in the $xy$-plane, using the polar angle $\theta$, with $\mathbf{k}=k(\cos{\theta},\sin{\theta},0)$. At an angle $\theta=\pi/2$, the double sum $S_\mathbf{k}$ simplifies into $S_\mathbf{0}=\sum_{j=1}^{N}\sum_{s\neq j}^{N}\cos{(\phi_{s}-\phi_{j})}$. For $S_\mathbf{0}=0$, spin-squeezing inequalities do not capture entanglement. Focusing on states with phases $\phi_n=n \delta$, this condition is reached when $\delta$ is a solution of $T_N(\cos\delta)-N\cos\delta+(N-1)=0$, with $T_N$ the Chebyshev polynomial of the first kind~\cite{SM}. The angular dependence of the witness $W_\mathbf{k}$ for such a case is shown in Fig.~\SubFig{Fig:Decay}{(a)}: The witness is negative for any direction of observation, thus detecting entanglement, except for $\theta=\pi/2$ where spin-squeezing witness $W_\mathbf{0}$ is measured. We point out that the detection of these single-excitation states by the electric field witness is valid for arbitrary particle numbers.

{\it Dynamical generation of entanglement.---}Beyond the issue of entanglement detection, let us now discuss the generation of entanglement via collective spontaneous emission. We consider two-level atoms driven resonantly, whose interaction through the vacuum modes results in an effective dipole-dipole interaction between them. In three dimensions, this interaction can be described by the following master equation~\cite{Lehmberg_1970,Lehmberg_1970_2}
\begin{align}
     \frac{d\hat{\rho}}{dt}&=\sum^{N}_{j,m\neq j}i\Delta_{jm}[\hat{\sigma}^{+}_{j}\hat{\sigma}^{-}_{m},\hat{\rho}]+ \mathcal{L}(\hat{\rho}), 
\end{align}
where we have set $\hbar\equiv 1$. The first part represents the coherent component of the dipole-dipole interaction, while the second term corresponds to the dissipative part, ruled by the Lindbladian $\mathcal{L}(\hat{\rho})=\sum^{N}_{j,m}\Gamma_{jm}\left(\hat{\sigma}^{-}_{j}\hat{\rho}\hat{\sigma}^{+}_{m}-\frac{1}{2}\{\hat{\sigma}^{+}_{m}\hat{\sigma}^{-}_{j},\hat{\rho}\}\right)$. The excitation-exchange term $\Delta_{jm}$ and crossed decay rates $\Gamma_{jm}$ are given by the Green's tensor, $\Delta_{jm}\equiv -\hat{\epsilon}_{j}^{*}\cdot \R\{\mathbf{G}(\vec{r}_{jm})\}\cdot\hat{\epsilon}_{m} $ and $\Gamma_{jm}\equiv \hat{\epsilon}_{j}^{*}\cdot \I\{\mathbf{G}(\vec{r}_{jm})\}\cdot\hat{\epsilon}_{m} $, with $\hat{\epsilon}_{l}$ is the polarization of the $l$-th dipole, and $\vec{r}_{jm}=\vec{r}_{j}-\vec{r}_{m}$ the relative position of atoms $j$ and $m$.  In free space, the tensor for dipoles with transition frequency $\omega=kc=2\pi c/\lambda$ and linewidth $\Gamma$
is given by $\mathbf{G}(\vec{r}_{jm})=\frac{3\Gamma}{4}\frac{e^{ik r_{jm}}}{(kr_{jm})^{3}}[(k^{2}r^{2}_{jm}+ikr_{jm}-1)\1_{3}-(k^{2}r^{2}_{jm}+i3kr_{jm}-3)\frac{\vec{r}_{jm}\vec{r}^{T}_{jm}}{r^{2}_{jm}}]$ for $j\neq m$, and $\mathbf{G}(\vec{r}_{jj})=i\frac{\Gamma}{2}\1_{3}$.

The superradiant cascade from interacting dipoles prepared in an initially fully-excited state was first explored by Dicke~\cite{Dicke1954} using collective states with a given photon number, which are therefore entangled. Yet, later studies showed that a semi-classical approach described properly the accelerated radiance~\cite{Arecchi1972,MacGillivray1976,Gross1982}, and entanglement is not generated in that process~\cite{Wolfe2014}. Differently, the long-lived (subradiant) modes that eventually emerge~\cite{Guerin2016} become very close to mixtures of ground and collective single-excitation states, and thus entangled~\cite{Santos2022}. While the fully-excited state decays mostly toward the symmetric (superradiant) states, subradiant states are most efficiently generated by sending the system to a statistical mixture~\cite{Cipris2021, Santos2022}. This scenario is confirmed in our simulations by analyzing the electric field and its fluctuations. As illustrated in Fig.~\SubFig{Fig:Decay}{(b)} for the regular chain, entanglement is detected at a later time for an initially fully excited system ($\Gamma t\approx 3$), as compared to an initially maximally mixed state $\hat{\rho}=\1/2^{N}$ ($\Gamma t\approx 1$): This observation is consistent with entanglement being generated only at late times, and being generated more efficiently from a statistical mixture than from a fully-excited state. 

As an alternate probe for the presence of entanglement, let us now monitor the global (pairwise) concurrence, defined as $\mathcal{C}_{\text{glob}}=\sum_{j,s\neq j}\mathcal{C}(\hat{\rho}_{js})/N(N-1)$, where $\mathcal{C}(\hat{\rho}_{js})$ corresponds to the pairwise concurrence for the pair $(j,s)$, derived from its reduced density matrix $\hat{\rho}_{js}$~\cite{Hill_1997, Wootters_98}. As observed in Fig.~\SubFig{Fig:Decay}{(b)}, the concurrence becomes finite at a later time than the entanglement detected through the field fluctuations. The electric-field-based detection may thus be a particularly promising tool to study critical systems, where concurrence has been used to detect phase transitions~\cite{Osterloh_2002,Osborne_2002}.

{\it Three-dimensional clouds.---}While the one-dimensional chain offers access to a direct measurement of spin squeezing $W_\mathbf{0}$ through the electric field in directions orthogonal to the chain, three-dimensional clouds do not possess a symmetry that allows for this feature. Let us now demonstrate how the family of witnesses based on the electric field extends the detection of entanglement for these systems as well. We now consider a disordered cloud of $N=8$ two-level atoms in three dimensions, prepared in the antisymmetric product state $\ket{A}=\ket{+-+-+-+-}$, where we have introduced the single-atom state $\ket{\pm}=(\ket{\uparrow}\pm\ket{\downarrow})/\sqrt{2}$. Note that while the atoms, and thus the antisymmetric state, are ordered by positions in the linear chain, for the three-dimensional cloud the ordering in $\ket{A}$ is arbitrary. The evolution of the witness $W_\mathbf{k}$ during the decay dynamics is presented in Fig.~\SubFig{Fig:Decay}{(c)}: Similar features for the 3D (gray curve) and 1D (purple curve) configurations are observed, with entanglement being detected at an early time ($\Gamma t\approx 10^{-1}$) in both cases. This suggests that the electric field witness is equally efficient to probe entanglement in arbitrary geometries.

The emergence of entanglement on this short (superradiant) time scale can be understood as follows: The antisymmetric separable state $\ket{A}$ has a strong projection on both superradiant and subradiant entangled states. Due to the short distance between the atomic dipoles, the strong Hamiltonian interactions are responsible for shifting their relative phases on a short timescale, thus sending the system toward an entangled state. This is akin to the case of emitters with different energies~\cite{Hettich2002, Wang2010, Trebbia2022,Oliveira2023}, but here with a shift induced by dipole-dipole interactions. 

Resorting to collective observables is all the more critical when the size of the quantum system increases, since tomography, even if local measurements are available~\cite{Sherson2010,Christakis2023}, is no longer possible due to the size of the Hilbert space. The ability of such observables to capture the entanglement may nonetheless be affected by the system size since the number of remote, weakly interacting pairs of particles will be relatively larger. To investigate how the entanglement detection scales with the system size, we resort to the second-order cumulant approximation, where three-point and higher correlations are factorized in terms of one- and two-point ones~\cite{Helmut_2015}. The regular atomic chain is initially in state $\ket{A}$, from which it evolves in the presence of dipole-dipole interactions. The evolution of $W_\mathbf{k}$ reveals that entanglement is detected at earlier times as the lattice spacing is reduced, and as the system size $N$ is increased, see Fig.~\SubFig{Fig:Decay}{(d)}. While the former effect is quite intuitive, since the interactions become stronger at shorter inter-particle distances, the latter is rather a signature of long-range interactions. But this observation also demonstrates the ability of the field-based witness to detect entanglement efficiently in large quantum systems.

{\it Conclusion.---}In this work, we have discussed how measuring the electric field quadratures, the total fluorescence, and their fluctuations can serve to detect entanglement between the quantum emitters. To this end, we used a witness related to spin squeezing which incorporates the phases present in the electric field. This witness captures, in particular, single-excitation collective states, in almost any direction of observation. Derived for systems of qubits described by the Pauli group, the witness is thus also valid for ensembles with inhomogeneous broadening, a common case for artificial qubits such as nitrogen-vacancy centers~\cite{2021_Pezzagna_Meijer_apr}. An interesting prospect is to generalize these electric-field-based inequalities to quantum emitters with arbitrary spin~\cite{2014_Vitagliano_Toth_pra}, offering tools to probe multilevel entanglement in systems with qudits~\cite{2018_Kraft_Guhne_prl}. 

While finding an optimal witness remains an outstanding challenge~\cite{Lewenstein_2000}, the proposed family of witnesses presents the strong advantage of not relying on state tomography, but rather on measuring collective observables which will be particularly practical in some experiments. Note that the inversion population operator $\hat{Z}$, and its fluctuations, can be measured from the evolution of the fluorescence in an integrating sphere, for example --- the radiated power corresponds to a decrease of the excited population, or by probing the atomic populations using an auxiliary ancilla state.

The electric field operators in different directions actually correspond to specific relative phases between the atomic operators, set by the term $\mathbf{k}$ in Eq.~\eqref{eq:far-field}. While this corresponds to an already infinite family of witnesses since $\mathbf{k}$ spans the $4\pi$ solid angles, a much broader family of collective operators and associated witnesses can be introduced, by setting arbitrary relative phases between the atomic operators. This ensemble of witnesses, which would span a parameter space $[0; 2\pi]^{N-1}$ (one of the phases can be set arbitrarily) is much broader than the one based on the electric field operators, yet accessing it may bring back the requirement of challenging local measurements.

\begin{acknowledgments}
{\it Acknowledgments.---}P.R., A.C.S., A.C., and R.B. acknowledge the financial support of the São Paulo Research Foundation (FAPESP) (Grants No. 2018/15554-5, 2019/13143-0, 2019/12842-2, 2022/06449-9, 2023/07463-8, 2022/12382-4, and 2023/03300-7), from the Brazilian CNPq (Conselho Nacional de Desenvolvimento Científico e Tecnológico), Grant No. 313632/2023-5. R.B. and R.K. received support from the project STIC-AmSud (Ph879-17/CAPES 88887.521971/2020-00), CAPES-COFECUB (CAPES 88887.711967/2022-00). R.K. acknowledges the support by QuantERA ERA-NET Cofund in Quantum Technologies (GA No. 731473), project PACE-IN, and the European project ANDLICA, ERC Advanced grant No. 832219. A.C.S acknowledges the support by the Proyecto Sinérgico CAM 2020 Y2020/TCS-6545 (NanoQuCo-CM) from the Comunidad de Madrid.
\end{acknowledgments}


%



\onecolumngrid
\newpage

\begin{center}
{\large{ {\bf Supplemental Material for: \\ Detecting entanglement from macroscopic measurements of the electric field and its fluctuations}}}

\vskip0.5\baselineskip{Pedro Rosario~\orcidlink{0000-0002-7628-7373},$^{1}$ Alan C. Santos~\orcidlink{0000-0002-6989-7958},$^{1,2}$ Nicola Piovella,$^{3}$ Robin Kaiser,$^{4}$ Andr\'e Cidrim~\orcidlink{0000-0003-0007-2330},$^{1}$ and Romain Bachelard~\orcidlink{0000-0002-6026-509X}$^{1}$}

\vskip0.5\baselineskip{ {\it $^{1}$Departamento de Física, Universidade Federal de São Carlos,\\ Rodovia Washington Luís, km 235 - SP-310, 13565-905 São Carlos, SP, Brazil\\
$^{2}$Instituto de Física Fundamental (IFF), Consejo Superior de Investigaciones Científicas (CSIC), \\ Calle Serrano 113b, 28006 Madrid, Spain\\
$^{3}$Dipartimento di Fisica Aldo Pontremoli, Universit\`a degli Studi di Milano, Via Celoria 16, I-20133 Milano, Italy \& INFN Sezione di Milano, Via Celoria 16, I-20133 Milano, Italy\\
$^{4}$Universit\'e C\^ote d'Azur, CNRS, Institut de Physique de Nice, 06560 Valbonne, France
}}

\end{center}

\appendix

\section*{Derivation  of the entanglement witnesses}\label{Appendix:First}
\subsection{Definitions and phase-dependent basis}

Let us introduce the following basis of Pauli operators, which incorporate optical-path terms:
\begin{align}    
\hat{\sigma}^{\mathbf{k},x}_{j}=e^{-i\mathbf{k}.\mathbf{r}_j}\hat{\sigma}^{-}_{j}+e^{i\mathbf{k}.\mathbf{r}_j}\hat{\sigma}^{+}_{j}, \ \ \hat{\sigma}^{\mathbf{k},y}_{j}=i(e^{i\mathbf{k}.\mathbf{r}_j}\hat{\sigma}^{+}_{j}-e^{-i\mathbf{k}.\mathbf{r}_j}\hat{\sigma}^{-}_{j}), \ \ 
\hat{\sigma}^{z}_{j}=\ket{\uparrow}\bra{\uparrow}_{j}-\ket{\downarrow}\bra{\downarrow}_{j},
\end{align}
where $\hat{\sigma}^{-}_{j}=\ketbra{\downarrow}{\uparrow}_j$ and $\hat{\sigma}^{+}_{j}=\ketbra{\uparrow}{\downarrow}_j$. Then, a one-qubit state can be represented by the density matrix
\begin{align}
    \hat{\rho}_{j}=\frac{1}{2}(\1+\mathbf{v}^{\mathbf{k}}_{j}\cdot \bm{\sigma}^{\mathbf{k}}_{j}),
\end{align}
where $\mathbf{v}^{\mathbf{k}}_{j}=(v^{\mathbf{k},x}_{j},v^{\mathbf{k},y}_{j},v^{z}_{j})$ is the coherence vector for atom $j$ and $\bm{\sigma}^{\mathbf{k}}_{j}=(\hat{\sigma}^{\mathbf{k},x}_{j},\hat{\sigma}^{\mathbf{k},y}_{j},\hat{\sigma}^{z}_{j})$. Any density matrix $\hat{\rho}$  satisfies $\text{Tr}(\hat{\rho}^{2})\leq 1$, which translates into
\begin{align}
(v^{\mathbf{k},x}_{j})^{2}+ (v^{\mathbf{k},y}_{j})^{2}+\left(v^{z}_{j}\right)^{2}\leq 1.
\label{eq:normalization}
\end{align}
The Bloch vector elements are defined as
\begin{align}
v^{\mathbf{k},\eta}_{j}=\text{Tr}\left[\hat{\sigma}^{\mathbf{k},\eta}_{j}\hat{\rho}_{j}\right], \ \ \text{with} \ \ \eta \ \in \{x,y\} \ \text{and}\ \ v^{z}_{j}=\text{Tr}\left[\hat{\sigma}^{z}_{j}\hat{\rho}_{j}\right].
    \label{eq:BV_definition}
\end{align}
Using the phase-dependent basis $\bm{\sigma}^{\mathbf{k}}_{j}$, an arbitrary one-qubit state can be written as
\begin{align}
   \hat{\rho}_{j} =\begin{pmatrix}
a& b+ic\\
b-ic & d
\end{pmatrix},
\ \ \ \text{with} \ \ |a|^{2}+|b|^{2}=1.
\end{align}

Inserting this expression into Eq. \eqref{eq:BV_definition} leads to
\begin{align}
    &v^{\mathbf{k},x}_{j}=2\left[b\cos(\mathbf{k}.\mathbf{r}_j)-c\sin(\mathbf{k}.\mathbf{r}_j)\right],\ \ \ v^{\mathbf{k},y}_{j}=-2\left[b\sin(\mathbf{k}.\mathbf{r}_j)+\cos(\mathbf{k}.\mathbf{r}_j)c\right], \ \ \ v^{z}_{j}=a-d,
\end{align}
so that $\mathbf{v}^{\mathbf{k}}_{j} \in \mathbb{R}^{3}$. 

We now consider the electric field radiated by the two-level atoms in the far field, $\hat{E}_\mathbf{k}^{\pm}=\sum_{j=1}^{N}e^{\mp i\mathbf{k}.\mathbf{r}_j}\hat{\sigma}^{\mp}_{j}$, and express its quadrature, $\hat{X}_{\mathbf{k}}$ and $\hat{Y}_{\mathbf{k}}$, and population imbalance, $\hat{Z}$, in this basis:
\begin{align}
&\hat{X}_{\mathbf{k}}=\hat{E}^{+}_{\mathbf{k}}+\hat{E}^{-}_{\mathbf{k}}=\sum_{j}^{N}\left(e^{ i\mathbf{k}.\mathbf{r}_j}\hat{\sigma}^{+}_{j}+e^{- i\mathbf{k}.\mathbf{r}_j}\hat{\sigma}^{-}_{j}\right)=\sum_{j=1}^{N}\hat{\sigma}^{\mathbf{k},x}_{j},\\
    &\hat{Y}_{\mathbf{k}}=i(\hat{E}^{+}_{\mathbf{k}}-\hat{E}^{-}_{\mathbf{k}})=\sum_{j}^{N}i\left(e^{ i\mathbf{k}.\mathbf{r}_j}\hat{\sigma}^{+}_{j}-e^{- i\mathbf{k}.\mathbf{r}_j}\hat{\sigma}^{-}_{j}\right)=\sum_{j=1}^{N}\hat{\sigma}^{\mathbf{k},y}_{j},\\
    &\hat{Z}=\sum_{j=1}^{N}\hat{\sigma}^{z}_{j}.
\end{align}

\subsection{First witness}
Let us first evaluate the quantity  $\langle \hat{X}^{2}_{\mathbf{k}} \rangle + \langle \hat{Y}^{2}_{\mathbf{k}} \rangle + \langle \hat{Z}^{2} \rangle$ for separable states, in order to determine an upper or lower bound in terms of the number of particles. A separable state with $N$ particles reads
\begin{align}
    \hat{\rho}=\sum_{l=1}^{L}p_{l}\hat\rho^{(l)}_{1}\otimes \hat\rho^{(l)}_{2}\otimes \hdots \otimes \hat\rho^{(l)}_{N}\ \ \ \text{with} \ \sum_{l=1}^{L}p_{l}=1.
\end{align}
We then take $\hat{\alpha} \in \{\hat{X},\hat{Y}\}$, which leads to
\begin{align}
\hat{\alpha}^{2}_{\mathbf{k}}=\sum_{j=1}^{N}\sum_{s=1}^{N}\hat{\sigma}^{\mathbf{k},\alpha}_{j} \hat{\sigma}^{\mathbf{k},\alpha}_{s}=\sum_{j=1}^{N}\1 +\sum_{j=1}^{N}\sum_{s\neq j}^{N}\hat{\sigma}^{\mathbf{k},\alpha}_{j} \hat{\sigma}^{\mathbf{k},\alpha}_{s}=N +\sum_{j=1}^{N}\sum_{s\neq j}^{N}\hat{\sigma}^{\mathbf{k},\alpha}_{j} \hat{\sigma}^{\mathbf{k},\alpha}_{s}.
\end{align}
Its expectation value then reads
\begin{align}
\langle \hat{\alpha}^{2}_{\mathbf{k}}\rangle &\nonumber= N+\text{Tr}\left[\left(\sum_{l=1}^{L}p_{l}\hat\rho^{(l)}_{1}\otimes \hat\rho^{(l)}_{2}\otimes \hdots \otimes \hat\rho^{(l)}_{N}\right)\left(\sum_{j=1}^{N}\sum_{s\neq j}^{N}\hat{\sigma}^{\mathbf{k},\alpha}_{j} \hat{\sigma}^{\mathbf{k},\alpha}_{s}\right)\right]\\
    &\nonumber=N +\sum_{j=1}^{N}\sum_{s\neq j}^{N}\sum_{l=1}^{L}p_{l}\text{Tr}\left[\hat\rho^{(l)}_{j}\hat{\sigma}^{\mathbf{k},\alpha}_{j}\right]\text{Tr}\left[\hat\rho^{(l)}_{s}\hat{\sigma}^{\mathbf{k},\alpha}_{s}\right]\\
    &=N+\sum_{j=1}^{N}\sum_{s\neq j}^{N}\sum_{l=1}^{L}p_{l}v^{\mathbf{k},\alpha,l}_{j}v^{\mathbf{k},\alpha,l}_{s}.
\end{align}
This allows us to write
\begin{align}
    \nonumber \langle \hat{X}^{2}_{\mathbf{k}}\rangle + \langle \hat{Y}^{2}_{\mathbf{k}}\rangle + \langle \hat{Z}^{2}\rangle =3N+\sum_{j=1}^{N}\sum_{s\neq j}^{N}\sum_{l=1}^{L}p_{l}(v^{\mathbf{k},x,l}_{j}v^{\mathbf{k},x,l}_{s}+v^{\mathbf{k},y,l}_{j}v^{\mathbf{k},y,l}_{s}+v^{z,l}_{j}v^{z,l}_{s})=3N+\sum_{j=1}^{N}\sum_{s\neq j}^{N}\sum_{l=1}^{L}p_{l}\mathbf{v}_{j}^{\mathbf{k},l}\cdot \mathbf{v}_{s}^{\mathbf{k},l}.
\end{align}
Here $\mathbf{v}_{j}^{\mathbf{k},l}\cdot \mathbf{v}_{s}^{\mathbf{k},l}$ stands for the dot product between two Bloch vectors. Therefore using Eq.~\eqref{eq:normalization} leads to $\mathbf{v}_{j}^{\mathbf{k},l}\cdot \mathbf{v}_{s}^{\mathbf{k},l}\leq 1$ and eventually to the first inequality:
\begin{align}
    \langle \hat{X}^{2}_{\mathbf{k}}\rangle + \langle \hat{Y}^{2}_{\mathbf{k}}\rangle + \langle \hat{Z}^{2}\rangle \leq  N(2+N) .
\end{align}
This results in the definition of the first witness~\eqref{ineq1}.

\subsection{Second witness}

Let us now evaluate the sum of the variances, $(\Delta \hat{X}_{\mathbf{k}})^{2}+(\Delta \hat{Y}_{\mathbf{k}})^{2}+(\Delta \hat{Z})^{2}$. 
The concavity of the variances makes that, for separable states:
\begin{align}
    (\Delta \hat{\alpha}_{\mathbf{k}})^{2}\geq \sum_{l=1}^{L}\sum_{j=1}^{N}p_{l}\left[\Delta \hat{\sigma}^{\mathbf{k},\alpha}_{j}\right]^{2}_{l}, \ \ \text{with}\ \ \hat{\alpha} \in \{\hat{X},\hat{Y}\},
    \label{eq:concavity}
\end{align}
where $\left[\Delta \hat{\sigma}^{\mathbf{k},\alpha}_{j}\right]^{2}_{l}$ is the (local) variance associated with particle $j$. This leads to
\begin{align}
    \nonumber(\Delta \hat{X}_{\mathbf{k}})^{2}+(\Delta \hat{Y}_{\mathbf{k}})^{2}+(\Delta \hat{Z})^{2}&\geq \sum_{l=1}^{L}\sum_{j=1}^{N}p_{l}\left(\left[\Delta (\hat{\sigma}^{\mathbf{k},x}_{j})\right]^{2}_{l}+\left[\Delta (\hat{\sigma}^{\mathbf{k},y}_{j})\right]^{2}_{l}+\left[\Delta (\hat{\sigma}^{z}_{j})\right]^{2}_{l}\right)\\
    &\nonumber=\sum_{l=1}^{L}\sum_{j=1}^{N}p_{l}\left(3-\left[\langle \hat{\sigma}^{\mathbf{k},x}_{j} \rangle^{2} \right]_{l} -\left[\langle \hat{\sigma}^{\mathbf{k},y}_{j}\rangle^{2} \right]_{l} -\left[\langle \hat{\sigma}^{z}_{j} \rangle^{2} \right]_{l} \right)\\
    &\nonumber=\sum_{l=1}^{L}\sum_{j=1}^{N}p_{l}\left(3-\left[ \left(v^{\mathbf{k},x}_{j}\right)^{2}\right]_{l} -\left[ \left(v^{\mathbf{k},y}_{j}\right)^{2}\right]_{l} -\left[\left(v^{z}_{j}\right)^{2}\right]_{l}\right)\\
    &\nonumber=\sum_{l=1}^{L}\sum_{j=1}^{N}p_{l}\left(3-\left[ \left(v^{\mathbf{k},x}_{j}\right)^{2}+\left(v^{\mathbf{k},y}_{j}\right)^{2}+\left(v^{z}_{j}\right)^{2}\right]_{l}\right)\\
    &\nonumber=2\sum_{l=1}^{L}\sum_{j=1}^{N}p_{l}=2N,
\end{align}
where we have used that $(v^{\mathbf{k},x}_{j})^{2}+(v^{\mathbf{k},y}_{j})^{2}+\left(v^{z}_{j}\right)^{2}=1$ for pure states to set the lower bound. Hence, we obtain the inequality from which the second witness, Eq.~\eqref{ineq2}, is derived:
\begin{align}
   (\Delta \hat{X}_{\mathbf{k}})^{2}+(\Delta \hat{Y}_{\mathbf{k}})^{2}+(\Delta \hat{Z})^{2} \geq 2N.
\end{align}

\subsection{Third witness}
The third inequality is obtained by analyzing $(N-1)(\Delta \hat{A})^{2}-\langle \hat{B}^{2} \rangle -\langle \hat{C}^{2} \rangle $, where $(\hat{A},\hat{B},\hat{C})$ is any cyclic permutation of $(\hat{X}_{\mathbf{k}},\hat{Y}_{\mathbf{k}},\hat{Z})$. Using the above relations for $\hat{\alpha},\hat{\beta}\in \{\hat{X},\hat{Y}\}$ we get
\begin{align}
    &(\Delta \hat{\alpha}_{\mathbf{k}})^{2}=N +\sum_{j=1}^{N}\sum_{s\neq j}^{N}\langle\hat{\sigma}^{\mathbf{k},\alpha}_{j} \hat{\sigma}^{\mathbf{k},\alpha}_{s}\rangle-\sum_{j=1}^{N}\sum_{s=1}^{N}\langle\hat{\sigma}^{\mathbf{k},\alpha}_{j}\rangle \langle\hat{\sigma}^{\mathbf{k},\alpha}_{s}\rangle, \\
    &  \langle \hat{\beta}^{2}_{\mathbf{k}} \rangle=N +\sum_{j=1}^{N}\sum_{s\neq j}^{N}\langle\hat{\sigma}^{\mathbf{k},\beta}_{j} \hat{\sigma}^{\mathbf{k},\beta}_{s}\rangle.
\end{align}
Taking into account the concavity relation in Eq.\eqref{eq:concavity} and the inequality 
\begin{align}
-\left(\sum_{j=1}^{N}v^{\mathbf{k},\alpha}_{j}\right)^{2}\geq -N\sum_{j=1}^{N}(v^{\mathbf{k},\alpha}_{j})^{2},
\label{eq:Ineq_BV_Elements}
\end{align}
we obtain
\begin{align}
    (N-1)(\Delta \hat{A})^{2}-\langle \hat{B}^{2} \rangle -\langle \hat{C}^{2} \rangle \geq -2N,
\end{align}
from which entanglement witness Eq.~\eqref{ineq3} is obtained.

\subsection{Fourth witness}
The last set of witnesses rely on the quantity $(N-1)\left[(\Delta \hat{A})^{2}+(\Delta \hat{B})^{2}\right]-\langle \hat{C}^{2} \rangle $, where $(\hat{A},\hat{B},\hat{C})$ are cyclic permutations of $(\hat{X}_{\mathbf{k}},\hat{Y}_{\mathbf{k}},\hat{Z})$. Using the concavity argument in Eq.\eqref{eq:concavity}, combined with Eq.\eqref{eq:Ineq_BV_Elements}, we get
\begin{align}
    (N-1)\left[(\Delta \hat{A})^{2}+(\Delta \hat{B})^{2}\right]-\langle \hat{C}^{2} \rangle  \geq N (N-2).
\end{align}
\section*{Single excitation collective states}\label{Appendix:second}
\subsection{Dicke state}\label{Appendix:second_1}
We here derive the properties for single excitation Dicke state $\hat{\rho }$, written under the form 
\begin{align}
    \hat{\rho }=\ketbra{\mathcal{D}(N)}{\mathcal{D}(N)}, \ \text{ with } \mathcal{D}(N)=\sum_{n}\frac{e^{i\phi_{n}}}{\sqrt{N}}\ket{\uparrow_{n}}.
\end{align}
The expectation value $\langle \hat{Y}^{2}_{\mathbf{k}} \rangle $ takes the form
\begin{align}
    \nonumber\langle \hat{Y}^{2}_{\mathbf{k}}\rangle &=\sum_{j=1}^{N}\sum_{s=1}^{N}\langle\hat{\sigma}^{\mathbf{k},y}_{j}\hat{\sigma}^{\mathbf{k},y}_{s}\rangle\\
    &\nonumber=N+\sum_{j=1}^{N}\sum_{s\neq j}^{N}\langle\hat{\sigma}^{\mathbf{k},y}_{j}\hat{\sigma}^{\mathbf{k},y}_{s}\rangle\\
    &\nonumber=N+\text{Tr}\left[\ketbra{\mathcal{D}(N)}{\mathcal{D}(N)}\left(\sum_{j=1}^{N}\sum_{s\neq j}^{N}\hat{\sigma}^{\mathbf{k},y}_{j}\hat{\sigma}^{\mathbf{k},y}_{s}\right)\right]\\
    &\nonumber=N+\sum_{j=1}^{N}\sum_{s\neq j}^{N}\bra{\mathcal{D}(N)}\hat{\sigma}^{\mathbf{k},y}_{j}\hat{\sigma}^{\mathbf{k},y}_{s}\ket{\mathcal{D}(N)}.
\end{align}
Using the properties $\hat{\sigma}^{\mathbf{k},y}_{s}\ket{\downarrow}=ie^{i\mathbf{k} \cdot \mathbf{r}_{s}}\ket{\uparrow}$, $\hat{\sigma}^{\mathbf{k},y}_{s}\ket{\uparrow}=-ie^{-i\mathbf{k} \cdot \mathbf{r}_{s}}\ket{\downarrow}$, we obtain
\begin{align}
    \nonumber\bra{\mathcal{D}(N)}\hat{\sigma}^{\mathbf{k},y}_{j}\hat{\sigma}^{\mathbf{k},y}_{s}\ket{\mathcal{D}(N)}&=\frac{1}{N}\sum_{m}\sum_{n}e^{i(\phi_{n}-\phi_{m})}(\delta_{sn}\delta_{jn}\delta_{mn}e^{i\mathbf{k}\cdot (\mathbf{r}_{j}-\mathbf{r}_{s})}+\delta_{sn}(1-\delta_{jn})\delta_{mj}e^{i\mathbf{k}\cdot (\mathbf{r}_{j}-\mathbf{r}_{s})}),\\
   &+\frac{1}{N}\sum_{m}\sum_{n}e^{i(\phi_{n}-\phi_{m})}((1-\delta_{sn})\delta_{jn}\delta_{ms}e^{-i\mathbf{k}\cdot (\mathbf{r}_{j}-\mathbf{r}_{s})}+(1-\delta_{sn})\delta_{js}\delta_{mn}e^{-i\mathbf{k}\cdot (\mathbf{r}_{j}-\mathbf{r}_{s})}),
\end{align}
which then leads to
\begin{align}
    \nonumber\sum_{j=1}^{N}\sum_{s\neq j}^{N}\bra{\mathcal{D}(N)}\hat{\sigma}^{\mathbf{k},y}_{j}\hat{\sigma}^{\mathbf{k},y}_{s}\ket{\mathcal{D}(N)}&=\frac{1}{N}\sum_{j=1}^{N}\sum_{s\neq j}^{N}\sum_{m}\sum_{n}e^{i(\phi_{n}-\phi_{m})}(\delta_{sn}\delta_{mj}e^{i\mathbf{k}\cdot (\mathbf{r}_{j}-\mathbf{r}_{s})}+\delta_{jn}\delta_{ms}e^{-i\mathbf{k}\cdot (\mathbf{r}_{j}-\mathbf{r}_{s})})\\
    &\nonumber=\frac{1}{N}\sum_{j=1}^{N}\sum_{s\neq j}^{N}(e^{i(\phi_{s}-\phi_{j})}e^{i\mathbf{k}\cdot (\mathbf{r}_{j}-\mathbf{r}_{s})}+e^{i(\phi_{j}-\phi_{s})}e^{-i\mathbf{k}\cdot (\mathbf{r}_{j}-\mathbf{r}_{s})})\\
    &=\frac{2}{N}\sum_{j=1}^{N}\sum_{s\neq j}^{N}\cos{\left[(\phi_{s}-\phi_{j})+\mathbf{k}\cdot(\mathbf{r}_{j}-\mathbf{r}_{s})\right]}.
\end{align}
It is straightforward to verify that $\langle \hat{Y}^{2}_{\mathbf{k}} \rangle=\langle \hat{X}^{2}_{\mathbf{k}} \rangle$, so  we finally obtain
\begin{align}
    &\langle \hat{X}^{2}_{\mathbf{k}} \rangle=N+\frac{2}{N}\sum_{j=1}^{N}\sum_{s\neq j}^{N}\cos{\left[(\phi_{s}-\phi_{j})+\mathbf{k}\cdot(\mathbf{r}_{j}-\mathbf{r}_{s})\right]},\\
    &\langle \hat{Y}^{2}_{\mathbf{k}} \rangle=N+\frac{2}{N}\sum_{j=1}^{N}\sum_{s\neq j}^{N}\cos{\left[(\phi_{s}-\phi_{j})+\mathbf{k}\cdot(\mathbf{r}_{j}-\mathbf{r}_{s})\right]},\\
    &\langle\hat{Z}^{2} \rangle = N+(N-4)(N-1),\\
    &\langle\hat{Z} \rangle = (N-2),\\
    &\langle\hat{Y}_{\mathbf{k}} \rangle =0,\\
    &\langle\hat{X}_{\mathbf{k}} \rangle = 0.
\end{align}
Note that if one omits the phases related to the field (or sets them to zero), the second and third inequalities derived previously take the form
\begin{align}
  & 0  \leq \frac{4}{N}\sum_{j=1}^{N}\sum_{s\neq j}^{N}\cos{(\phi_{j}-\phi_{s})},
  \label{eq:11}\\
   & 0  \leq -\frac{4}{N}\sum_{j=1}^{N}\sum_{s\neq j}^{N}\cos{(\phi_{j}-\phi_{s})}.
  \label{eq:32}
\end{align}
Hence, unless the single-excitation Dicke state has phases which satisfy $\sum_{j=1}^{N}\sum_{s\neq j}^{N}\cos{(\phi_{j}-\phi_{s})}=0$, either \eqref{eq:11} or \eqref{eq:32} is violated, and the entanglement is detected.

If the optical path terms are considered, one obtains the full inequalities
\begin{align}
    &\nonumber 0\leq \frac{4}{N}\sum_{j=1}^{N}\sum_{s\neq j}^{N}\cos{\left[(\phi_{s}-\phi_{j})+\mathbf{k}\cdot(\mathbf{r}_{j}-\mathbf{r}_{s})\right]},\\
    &0\leq -\frac{4}{N}\sum_{j=1}^{N}\sum_{s\neq j}^{N}\cos{\left[(\phi_{s}-\phi_{j})+\mathbf{k}\cdot(\mathbf{r}_{j}-\mathbf{r}_{s})\right]}.
    \label{eq:B12}
\end{align}
In particular, changing the direction of observation alters the phase terms in the cosine, so while the inequality may not be violated for a specific angle, it should then be violated for a different one.

If we now look for a specific set of phases of the form $\{\phi_{n}=n\nu\}_{n=1}^{N}$ which satisfy the relation
\begin{align}
    \sum_{j=1}^{N}\sum_{s\neq j}^{N}\cos{(\phi_{j}-\phi_{s})}=\frac{\cos(N\nu)-1}{\cos(\nu)-1}-N=0,
\end{align}
we obtain
\begin{align}\label{eq:poly}
    \frac{\cos(N\nu)-1}{\cos(\nu)-1}-N=N\delta -T_{N}(\delta)+(1-N)=0,
\end{align}
where
$T_{N}(\delta)$ are the Chebyshev polynomials of the first kind, and where $\delta$ should be different from -1 and 1. Then the set of phases takes the form 
\begin{align}
    \{\phi_{n}\}_{n=1}^{N}=\{n\arccos{\delta}\}_{n=1}^{N},
\end{align}
with $\delta$ a solution of \eqref{eq:poly}. For the specific case of $N=2$ qubits, the solution is $\delta=0$, which leads to the phases
\begin{align}
    \{\phi_{n}\}_{n=1}^{2}=\left\{0,\frac{\pi}{2}\right\},
\end{align}
up to a global phase.

\end{document}